# Vacancy-Mediated Disordering Process in Binary Alloys at Finite Temperatures: Monte Carlo Simulations

J. Wong-ekkabut,* W. Triampo and I-Ming Tang

*Department of Physics, Faculty of Science, Mahidol University, Bangkok 10400, Thailand and*
*Capability Building Unit in Nanoscience and Nanotechnology,*
*Faculty of Science, Mahidol University, Bangkok 10400, Thailand*

D. Triampo

*Department of Chemistry, Faculty of Science, Mahidol University, Bangkok 10400, Thailand*

D. Baowan and Y. Lenbury

*Department of Mathematics, Faculty of Science, Mahidol University, Bangkok 10400, Thailand*

We have investigated the time evolution of the vacancy-mediated disordering process in binary alloys at finite temperatures. Qualitatively, we monitor the changes in the configurations by taking sequences of snapshots for various temperatures and comparing their morphologies. Quantitatively, we carry out Monte Carlo simulations to determine the time-dependent disorder parameter $A(L, T; t)$ and the time-dependent structure factors $S_{\bm{k}}(t)$ for moderately low temperatures. This study differs from previous studies done at infinite temperature in that the vacancy here executes highly active walks, which are subjected to nonlinear feedbacks, instead of the random walks that take place in the limit of $T \to \infty$. We find that the slope of the log-log plot of $A(L, T; t)$ vs. $t$ for finite temperatures follows the temperature dependence given by the Gompertz function and reaches a limiting value of $\frac{1}{2}$ only as the temperature approaches infinity. For the structure factors, namely, $S_{\bm{k}}(t)$ vs. $t$, the overall features are similar to those found at infinite $T$. The two key differences between our results and those at infinite $T$ are the saturated value and the intermediate region in which the portion of the graph whose slope is equal to one becomes smaller and is gradually replaced by a curve having a slope of 0.5. This last difference is especially evident at very low temperatures.

## I. INTRODUCTION

Vacancy-mediated disordering processes play very important roles in material science and physics. In particular, they are important in interface corrosion or erosion phenomena [1–3] and in device fabrication [4, 5]. We are interested here in how disorder develops in a system when its kinetics is controlled by an energetic "feedback" from the environment. We have investigated the kinetic disordering caused by a highly mobile vacancy arising when there is a rapid increase in the temperature of an initially phase-segregated binary alloy. We start with a ferromagnetically ordered configuration with sharp interfaces at zero-temperature and end up with a disordered configuration at a final temperature [see Fig. 1]. We are interested in the evolution of how particles of one species are transported into regions that are dominated by other species when there is interface roughening or when a complete destruction of the interface has occurred. If the final temperature is sufficiently high, the interface will eventually disappear, resulting in a homogeneous final (mixed) state. If the temperatures are low, the "degree of homogeneity" associated with the final configuration will change.

We will be looking at a symmetric binary alloy of A and B atoms containing a very small number of vacancies, $\approx 10^{-5}$, the number occurring in most real systems [6]. The vacancy acts as a "catalyst", exchanging with neighboring particles according to the usual energetic of the (dilute) Ising model [7]. The particle themselves form a passive background whose dynamics is slaved to the vacancy motion. This system corresponds to a real material in which the characteristic time scale for vacancy diffusion is much faster than the ordinary bulk

---

*E-mail: jirasakwo@hotmail.com

diffusion time.

The complete or partial mixing of two materials at the interface plays a key role in many physical processes. We mention just one application having huge technological potential for device fabrication. It concerns nanowire etching by electron beam lithography [4]: When a thin film of platinum is deposited on a silicon wafer, inter-diffusion of $Pt$ and $Si$ will produce a mixed layer. If this layer is heated locally, *e.g.*, by exposure to an electron beam, silicides, such as $Pt_2Si$ and $PtSi$, will be formed. The unexposed platinum can then be etched away, leaving behind conducting nanoscale structures. The performance of these devices requires precisely engineered layer thicknesses and interfaces. The performance can be significantly changed by the disorder arising from the inter-diffusion or the resulting interfacial fluctuations.

As was done in many studies on phase ordering, *e.g.*, on phase separation and domain growth [8–16], we will be looking for universal features in the time evolution of the system. Several questions emerge quite naturally: Are there characteristic time scales over which the disordering takes place, how do they depend on the system size, temperature, and other controlling parameters, how do local density profiles and correlation functions evolve with time, and how do these features respond to changes in the relative concentrations of vacancies and alloy components? Answers to some of these questions are known, but only in the limit of infinitely high temperatures where the vacancies execute random walk-motion [17]. In this paper we will study the "up quenching" to non-zero, moderate ($T > T_c$) and low($T < T_c$)temperatures. The extensions to these temperatures are nontrivial ones since the hopping rates of the vacancy will depend on the spin configuration in its vicinity. This will introduce a highly non-linear feedback effect, which is absent for the case of Brownian vacancies. If the final temperature remains above the coexistence line of the Ising model, the steady state will be disordered. The ordered phase will remain if the temperature is below $T_c$. Mathematically, there are no exact solutions to the non-equilibrium master equation [18] for this problem except for very small system [24]. Progress, therefore, relies mainly on computer simulations.

This article is organized as follows: First, we clearly define the model, the time-dependent disorder parameters, and the structure factors. This is done in Section II. In Section III, the Monte Carlo (MC) simulation results and discussion are given. We summarize our results and make some comments in the last section.

## II. MODEL

We start with a two-dimensional square lattice of dimensions $L \times L$. Each lattice site is denoted by a pair of integers, $\boldsymbol{r} = (i, j)$. To model binary systems consisting of two species of particles, we allow each site to be occupied by either a black particle ("spin up"), a white particle ("spin down"), or a vacancy ("spin zero"). Multiple occupancy is forbidden. One can view this model as being the dilute Ising model. A configuration in the model will be described by a set of spin variables $\{\sigma_{\boldsymbol{r}}\}$ which can take three values: $\sigma_{\boldsymbol{r}} = +1(-1)$ for a black (white) particle, and $\sigma_{\boldsymbol{r}} = 0$ for a vacancy. The numbers of black ($N^+$) and white ($N^-$) particles are conserved and differ by at most 1: $N^+ \approx N^- \cong \frac{1}{2}(L^2)$. To model the minute concentration of vacancies which is seen in real systems, we focus on the case where the number of vacancies is much less than the number of black and the number of white particles, $M \ll N^+, N^-$. In fact, we take $M$ to be 1 in all of the simulations. Fully periodic boundary conditions in all directions are assumed. The initial configuration will be completely phase segregated and unstable; *i.e.*, black and white particles will each fill one half of the system, with a sharp flat interface between them (chosen to lie horizontally along the $x$ axis). The single vacancy will initially be located at the interface.

We now turn to the microscopic dynamics or local update rule. Besides the correlation via the excluded volume constraint, the particles and the vacancy interact with one another according to the rules of the dilute Ising model:

$$H[\{\sigma_{\boldsymbol{r}}\}] = -J \sum_{<\boldsymbol{r},\boldsymbol{r}'>} \sigma_{\boldsymbol{r}}\sigma_{\boldsymbol{r}'} \qquad (1)$$

with ferromagnetic, nearest-neighbor coupling for $J > 0$. Since the dilution is very small, the behavior of the system will be that of the ordinary (non-dilute) two-dimensional Ising model. Therefore, a phase transition from a disordered phase to a phase-segregated phase will occur at the Onsager critical temperature, $T_c = 2.267...J/k_B$ [19]. The ground state will be doubly degenerate and consist of strips of positive and negative spins, each filling half of the system, separated by two planar interfaces. Since the particle-particle interactions are ferromagnetic, the single vacancy is at the interface. This is the initial configuration of the disordering process.

The vacancy now performs an active walk or non-linear feedback walk on the lattice with a transition rate per unit time step (Monte Carlo step) defined as $W[\{\sigma_{\boldsymbol{r}}\}\{\sigma'_{\boldsymbol{r}}\}]$ from configuration $\{\sigma_{\boldsymbol{r}}\}$ into a new configuration $\{\sigma'_{\boldsymbol{r}}\}$. Here, only particle-vacancy nearest-neighbor exchanges are allowed. We choose $W[\{\sigma_{\boldsymbol{r}}\}\{\sigma'_{\boldsymbol{r}}\}]$ to be the usual Metropolis rate [20]: namely,

$$W = min\{1, \exp(-1/k_BT)\Delta H)\} \qquad (2)$$

where $\Delta H$ is the difference in the energy of the system before and after the jump. One should expect that as time progresses, the vacancy will disorder the interface when $T > 0$. For $T > T_c > 0$, the vacancy will completely dissolve the interface.

In each Monte Carlo time step (MCS) in the simulations, one of the four nearest neighbors of the vacancy will be picked at random. The exchange with the picked neighbor site is then performed according to the rate defined above; no particle-particle exchanges are allowed. There are two control parameters for the simulations, namely the temperature $T$ and the system size $L$. The final temperature after the up-quench (measured in units of the Onsager temperature $T_c$) varies between $0.5T_c$ and infinity. The system size is $100 \times 100$. Our data are averages over $10^3$ realizations (or runs) or more, depending on the desired precision of the output.

At $t = 0$, a system which has the phase separation of the $T = 0$ configuration experiences a temperature up-quench to a finite temperature. We can now look at the evolution of the system as it undergoes disordering. We qualitatively monitor the temporal evolution visually and quantitatively measure the following:

### 1. Time-dependent Disorder Parameter $A(L,T;t)$

The average number of black and white nearest-neighbor pairs, $A(L,T;t)$, as a function of time is related to the Ising energy by

$$A(L,T;t) \cong L^2 + \frac{1}{2J} <H>, \qquad (3)$$

where $<\bullet>$ denotes the time-dependent configurational average over a large number of independent runs.

### 2. Time-dependent Structure Factors $S_\mathbf{k}(t)$

The timed-dependent structure facture factor $S_\mathbf{k}(t)$ is the Fourier transform of the correlation function defined as

$$S_\mathbf{k}(t) \equiv \frac{1}{L^2} < |\sum \sigma_\mathbf{r}(t) \exp(-i\mathbf{k} \cdot \mathbf{r})|^2 >, \qquad (4)$$

where $\mathbf{k} = (k_x, k_y) = \frac{2\pi}{L}(n_x, n_y)$, with $n_x$ and $n_y$ being integers. We shall focus on the more interesting structure factors, *i.e.*, those whose wave vector $\mathbf{k}$ is perpendicular to the initial planar interface. We shall focus on $n_y = 0$ and $n_x = 1, 2, 3, 4, 5,$ and $6$.

## III. RESULTS AND DISCUSSION

### 1. Visual Snapshots

Qualitatively, the visual impressions of the disordering process are seen in Fig. 1. For comparison, the visual impressions of the saturated configurations are seen in Fig. 2. The former shows the evolution of a typical configuration in a $100 \times 100$ system for $T = \infty$, and the later shows the saturated configuration for four temperatures ($T = 3.5T_c, 1.5T_c,$ and $0.8T_c$). In all cases, the initial interfaces at $t = 0$ were completely smooth. As time progressed, the interfaces began to break up slowly as the vacancy move: a random walk motion for $T = \infty$ and an active walk for finite temperatures. As more particles (inter) diffused into the oppositely colored domains, roughening of the interface began. As time went on, the degree of homogeneity became larger and the interface became rougher. Eventually, at steady state, the system corresponding $T > T_c$ to became disordered or became a homogenous mixed state. For $T < T_c$, phase segregation remains, but with a relatively rougher interface. As the final temperature was lowered farther, the system took a longer time to reach the final steady state. The time taken went approximately as the order of $L^4$ (for not too low of a temperature). Even from the snapshots, one can clearly see that in comparison with the infinite-temperature case, the final configurations for the non-zero temperatures show clear evidence of a finite correlation length (defined as the length scale associated with a two-point spatial correlation function), which will get larger as $T$ is set lower.

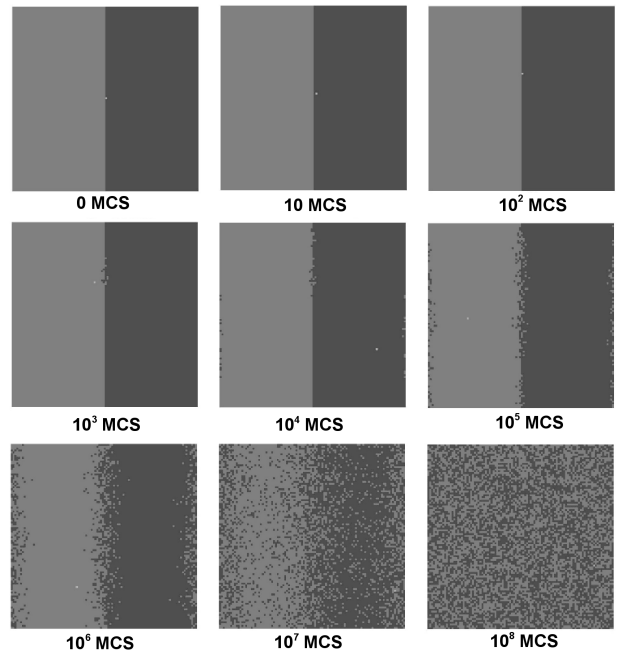

Fig. 1. Sequence of snapshots showing the disordering process of $100 \times 100$ a system with $T = \infty$. The black and the gray squares represent the two types of particles ($\sigma_\mathbf{r} = \pm 1$), and the white square denotes the vacancy ($\sigma_\mathbf{r} = 0$). The configurations were recorded after $0, 10, 10^2, 10^3, 10^4, 10^5, 10^6, 10^7,$ and $10^8$ MCS.

It should be noted that the "correlation" length measures the size of the fluctuation and diverges at criti-cal temperatures. One also sees larger clusters. These are due to the effects of the interactions between particles, which becomes more and more significant as T is decreased. It should be noted that, in all cases, the vacancy comes into equilibrium much sooner than the particles. At lower temperatures, it takes the vacancy longer to reach the steady state. From these visual pictures, it is evident that at least two time scales have to be taken into account: namely, the different time scales for the vacancy and the particles to reach the equilibrium state, separately [21–23] (vacancy time scale $\approx L^2$ and particles time scale $\approx L^4$ form random walk theory). The time scale for the vacancies can be determined by measuring the vacancy profile. This, however, would be extremely difficult for the case of highly correlated motions using the analytic method employed in this paper. It could be done by applying the two-time-scale technique to the exact master equation. Work on this new method is presently being done.

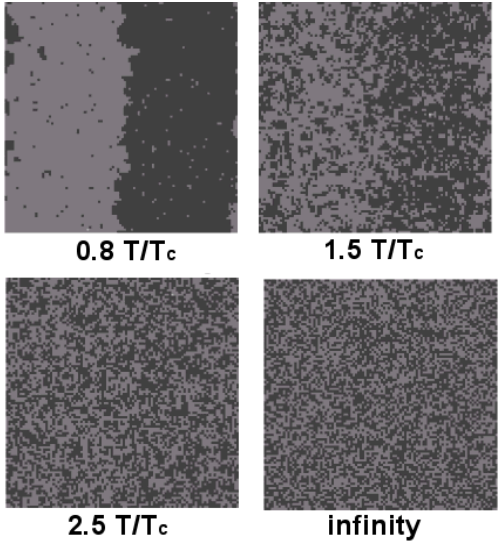

Fig. 2. Snapshots showing the typical saturated configurations of a $100 \times 100$ system in $10^8$MCS with $T = 0.8T_c, 1.5T_c, 2.5T_c$, and $\infty$.

## 2. Time-dependent Disorder Parameter

The disordering process is clearly reflected in the disorder parameter $A(L,T;t)$. As seen in Fig. 3 and Fig. 4, $A(L,T;t)$ increases from an order of $2L-1$ for the initial configuration to an order of $L^2$ for the equilibrium configuration. Similar to Refs. 21 and 22, one can clearly distinguish (at not so low a $T$), three regions in Fig. 3, an early region (E), an intermediate region (I), and a saturation region (S). It should be noted that the I and S regions emerge only in a finite system. As the system size increases, the (I) region spans a widening time range. In Ref. 21, it was shown that in I, the system shows a

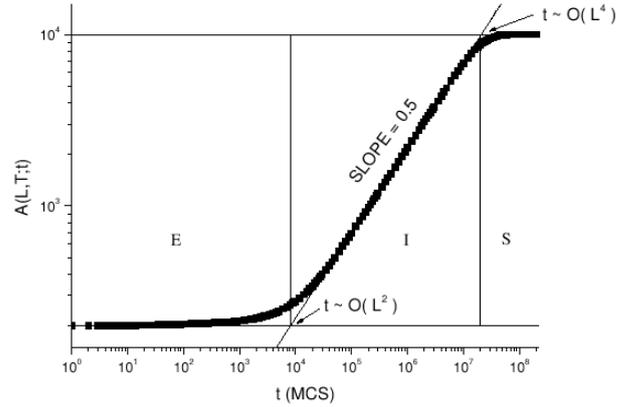

Fig. 3. Plot of the disorder parameter $A(L,T;t)$ vs. $t$ for $L = 100$ and $T = \infty$. It shows the emergence of an early region (E), an intermediate region (I), and a late or saturation region (S). The reference straight line has a slope of 0.5. The approximate saturation time scale of the vacancy and the particles are indicated on the bottom and the top, respectively.

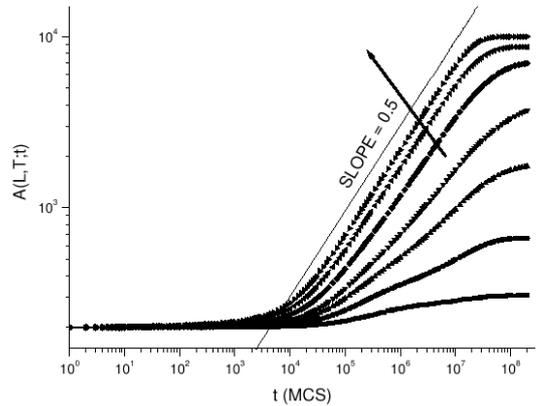

Fig. 4. Data for the disorder parameter $A(L,T;t)$ versus time at $0.5T_c, 0.7T_c, 0.9T_c, 1.1T_c, 1.7T_c, 3.5T_c$, and $\infty$ according to the arrow direction. We show the changing slope and the steady-state number of broken-bond lowering.

dynamic scaling as $A(L,T;t) \approx t^{\frac{1}{2}}$ for infinite $T$. The universality of this scaling relation has been extensively studied for the case of infinitely high $T$ with an external field [23]. The same scaling relation was found to remain valid in the case of high $T$ with an external field

One of the goals in this work was to test to what extend dynamic scaling, namely, the relation $A(L,T;t) \approx t^{\frac{1}{2}}$, holds at moderate and low temperatures. We carried out extensive MC simulations and fitted the curves using nonlinear regression. We found that the slope of the log-log plot or the power dependence (in $t$) of the disorder parameter at low temperature deviated significantly from the value 0.5. The results are seen in the Table1.

In Fig. 4, log-log plots $A(L,T;t)$ vs. $t$ for various finite temperatures are shown. These should be compared with the slope for the infinite-T case (whose slope = 0.5). In

Table 1. Slope of the intermediate region for a lattice size of which has a standard error of slope, a correlation coefficient (R), a standard deviation (SD), and a probability value. For each datum, we choose a range of time with 35 numbers from $10^5$ to $6 \times 10^6$.

| Temperature (T) | Slope | Standard error of slope | Correlation coefficient (R) | Standard deviation (SD) |
|---|---|---|---|---|
| 0.5 | 0.06460 | 0.00132 | 0.99322 | 0.00930 |
| 0.7 | 0.14225 | 8.44926E-4 | 0.99942 | 0.00601 |
| 0.9 | 0.25175 | 0.00153 | 0.99939 | 0.01090 |
| 1.0 | 0.30063 | 0.00198 | 0.99928 | 0.01411 |
| 1.1 | 0.34136 | 0.00226 | 0.99928 | 0.01608 |
| 1.3 | 0.39467 | 0.00228 | 0.99945 | 0.01623 |
| 1.5 | 0.42629 | 0.00182 | 0.99970 | 0.01297 |
| 1.7 | 0.44476 | 0.00160 | 0.99979 | 0.01136 |
| 1.9 | 0.45742 | 0.00122 | 0.99988 | 0.00866 |
| 2.1 | 0.46503 | 0.00105 | 0.99992 | 0.00748 |
| 2.3 | 0.47038 | 0.00083 | 0.99995 | 0.00594 |
| 2.5 | 0.47485 | 0.00074 | 0.99996 | 0.00524 |
| 2.7 | 0.47621 | 0.00072 | 0.99996 | 0.00514 |
| 2.9 | 0.47985 | 0.00030 | 0.99999 | 0.00211 |
| 3.1 | 0.48214 | 0.00048 | 0.99998 | 0.00340 |
| 3.3 | 0.48317 | 0.00042 | 0.99999 | 0.00296 |
| 3.5 | 0.48574 | 0.00029 | 0.99999 | 0.00208 |
| infinity | 0.49921 | 0.00012 | 1.00000 | 0.00088 |

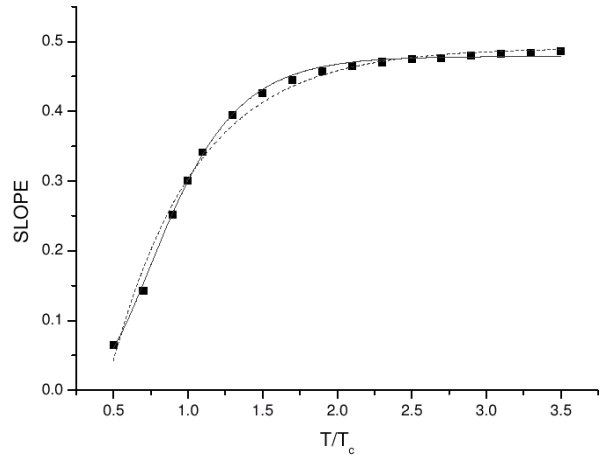

Fig. 5. Fitting of the first-order exponential function (dashed line) given in Eq. (5) with $y_0 = 0.49086 \pm 0.00587$, $A_1 = -1.07167 \pm 0.06719$, and $t_1 = 0.57345 \pm 0.0334$ with the regression $R^2 = 0.99072$ and $Chi^2 = 0.00018$, and the fitting of the Gompertz function (solid line) given in Eq. (6) with $a = 0.47867 \pm 0.00188$, $k = 2.98570 \pm 0.07369$, and $x_c = 0.74522 \pm 0.00604$ with the regression $R^2 = 0.99852$ and $Chi^2 = 0.00003$.

Table 1, the data analysis of the slope was done using the program Microcal(MT) Origin, version 6.0 [25]. For the intermediate region, we determined the slope by using a linear regression analysis of 35 collected points generated by $10^5$ to $6 \times 10^6$, more or less, Monte Carlo Steps (MCS) (~1.5 decades). From the table, we can see that at $T = 0.5T_c$, the slope is 0.0646 with a correlation coefficient = 0.99533 and a standard deviation (SD) = 0.0093. This deviates from 0.5 by about 87.08 %. At $T = 3.5T_c$, the slope is 0.48574 with a correlation coefficient = 0.99999 and a SD = 0.00208. This slope deviates from 0.5 by about 2.85 %. If the temperature is increased further to $T \gg T_c$ (infinite temperature), the slope increases to 0.5. As a function of T, the value of the slope increases rapidly at first and then slowly to its limiting value. We confirm this by fitting the data to several fitting functions. First, we used the first order exponential decay function

$$y = y_0 + A_1 e^{-\frac{x}{t_1}}, \qquad (5)$$

where $y_0$, $A_1$, and $t_1$ are fitting parameters. $t_1$ is relate to a temperature threshold. Next, the Gompertz function [26]

$$y = ae^{-(\exp(-k(x-x_c)))}, \qquad (6)$$

was used. Here $a$, $k$, and $x_c$ are fitting parameters. The Gompertz function has been used to analyze the dynamic growth of biological system, i.e., tumor growth. To simplify Eq. (6), one can do a Taylor expansion of the Gompertz function. The first two terms will be the first-order exponential decay terms, $y = a(1 - \exp(-k(x-x_c)) + \frac{\exp(-2k(x-x_c))}{2!} + ...)$. The fit of the data to Eq. (5) gives $y_0 = 0.49086 \pm 0.00587$, $A_1 = -1.07167 \pm 0.06719$, and $t_1 = 0.7345 \pm 0.0334$ with a regression $R^2 = 0.99072$ and $Chi^2 = 0.00018$. A fit of the data to the Gompertz functions gives $a = 0.47867 \pm 0.00188$, $k = 2.98570 \pm 0.07369$, and $x_c = 0.74522 \pm 0.00604$ with a regression $R^2 = 0.99852$ and $Chi^2 = 0.00003$, as shown in Fig. 5. From the values of $R^2$ and $Chi^2$, the Gompertz function appears to give a better fit to our data.

At not too low a $T$, this conclusion might be tenuous since one could argue that a correction to the dynamic scaling caused the deviation. After extensive simulations, we are convinced that the behavior at these not-too-low values of $T$'s cannot belong to the same universality class because of the great deviation from the slope 0.5. One explanation could be the following: In the case of the random walk or biased walk, the walker has no interaction with the environment. Therefore, there would be no feedback. The walker wanders through the system, regardless of the change in the environmental landscape. In contrast for the finite-T case, there is a non-linear feedback which causes the walker to become a more active walker. Changes in the landscape, the energy, or the barrier will cause a nonlinear feedback which will change how the walker takes its next step. The effect will be highly non-linear even though it is still controlled by a driving force which is forcing the system into the "appropriate" steady state. As the temperature decreases, the particle's motion becomes more and more correlated.

This results in the extinction of the random nature walk since it is driven by uncorrelated thermal fluctuations. We have a plot of the relation between the slope and the temperature in Fig. 5 (exponential function fit and Gompertz function fit)

With the significant change in the slope, it is reasonable to say that the scaling law for a highly active walk corresponding to moderate and low temperatures should be different from that for high temperatures or for a biased walk. It is obvious that their universality classes are not the same. For low temperatures, the microscopic dynamics is local. Due to the interaction between the particles, correlation effects dominate. As seen from the snapshots, the correlation length, $\zeta(T)$, depends on the temperature. As is well known, $\zeta(T)$ tends to $O(L)$ as $T$ approaches $T_c$. There is no characteristic length scale as in critical phenomena. The concept of a short distance over which there is no memory of the previous random walk breaks down. In our simulations, one notices that $\zeta(T)$ is in units of the lattice spacing.

## 3. Time-dependent Structure Factors (SF)

In this paper we give some preliminary, but significant, results for $S_{\bm{k}}(t)$. More simulations and analytic works were in progress. We focused here on the $S_{\bm{k}}(t)$'s whose wave vectors were perpendicular to the initial interfacial boundary, $n_y = 0$ and $n_x = 1, 2, 3, 4, 5$, and 6 to conserve computer run time. The remaining values of $n_x$ are expected to give features similar to those for the first six $n_x$'s without any dramatic change. We shall refer to them as the odd and the even wave-vector indices. From the definition of the SF's, $S_{\bm{k}}(t = 0)$ is on the order of $L^2$ when is $n_x$ odd and is on the order of $(L^2)^{-1}$ when is $n_x$ even. There would have been some changes in notation if we had used on odd lattice size. Since we have only considered $T > T_c$, the equilibrium configuration will be homogeneous. $S_{\bm{k}}(t \to \infty, T \gg 1) = 1$ for both odd and even wave vectors.

Some interesting results have been obtained in the infinite-temperature limit [27]. Like our results for finite temperatures, three regions emerge: namely, the early (E), the intermediate (I), and the saturation (S) regions, separated by two characteristic times, as shown in Fig. 6. $S_{\bm{k}}(t)$ starts off in the E region and remains there until a time on the order of $L^2$. This marks the onset of the I region. In this region, the odd SF's are governed by on exponential decay function. In contrast, the even SF's are governed by a power law with a slope of unity. They then develop surprising "dip"s before they finally reach the equilibrium region E. ($S_{\bm{k}}$ equal unity with a slope or about 0.5, as shown in the graph.)

Turning to the finite-T cases, we see that the three regions still exist. The time scales are longer than those for the infinite-temperature case because the correlations are much more complicated. Due to the different degrees of homogeneity at equilibrium, the saturation values or $S_{\bm{k}}(t \to \infty)$'s are different. They will be larger for lower $T$, as seen in Fig. 7. Interestingly, these equilibrium values vs. temperature can be fitted with an exponential function, as seen in Fig. 8. The dips seen before are still present, but become more and more obscured when the temperature is lowered. The portion of the graph exhibiting a slope scaling exponent of 1 in the 'I' regime now is replaced sooner by a curve with a slope = 0.5. This is clearly seen when the wave vectors are small, but is somewhat obscured when the wave vectors become large. This suggests that the high correlation due to the finite-T may break the universality of the scaling of the behavior at infinite $T$. This is caused by the microscopic mechanism of how vacancies move and how the non-linearities interact with the environment. This is not yet fully understood and is yet to be seen.

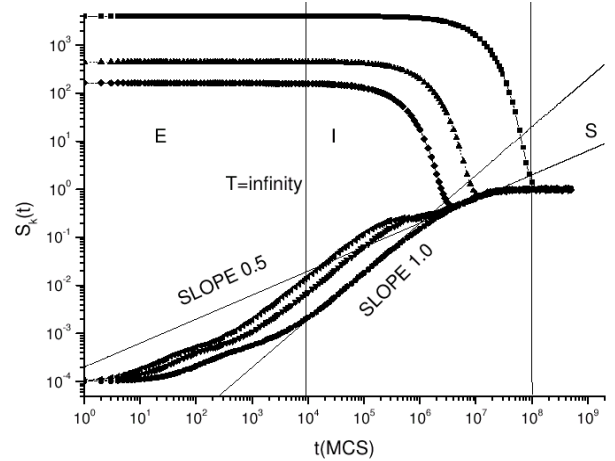

Fig. 6. Time evolution of the first six structure factors, $S_{\bm{k}}$, with $\bm{k} = (2\pi n/L, 0)$, $n_x = 1, 3, 5, 2, 4, 6$ from the top to the bottom, and a lattice size of 100. The straight lines have slopes 0.5 and 1.0 when the selected temperature is infinite.

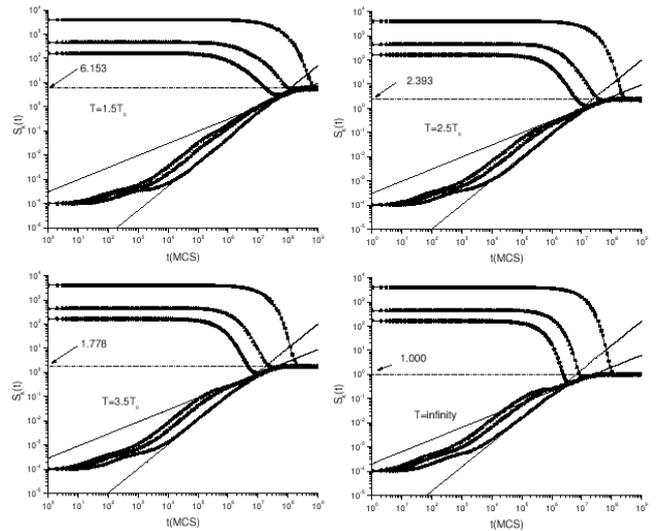

Fig. 7. Time evolution of the first six structure factors, $S_{\bm{k}}$, with $\bm{k} = (2\pi n/L, 0)$, $n_x = 1, 3, 5, 2, 4, 6$ from the top to the bottom, and a lattice size of 100. The straight lines have slopes 0.5 and 1.0 when the selected temperature are $1.5T_c, 2.5T_c, 3.5T_c$, and $\infty$.

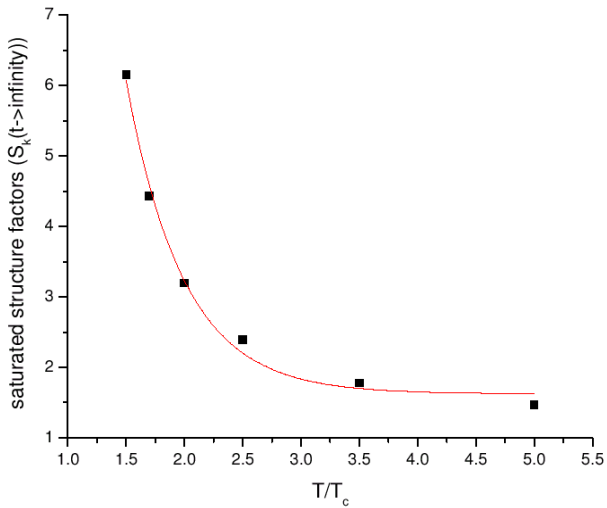

Fig. 8. Fitting of the saturated structure factor by using a first-order exponential function with $y_0 = 1.62458 \pm 0.13563$, $A_1 = 93.7220 \pm 33.50415$, and $t_1 = 0.49187 \pm 0.05733$ with the regression $R^2 = 0.99373$ and $Chi^2 = 0.03326$.

## IV. SUMMARY AND CONCLUSION

We have investigated the vacancy-driven disordering process in an initially phase-segregated binary system. We have considered up-quenches to finite temperatures, $T < \infty$, and compared the results with those found when the up-quenching was from $T = 0$ to $T = \infty$. Performing MC simulations for a range of temperatures, we sought to see whether the scaling exponents of the Brownian vacancy case still held. Qualitatively, we monitored the change of the configuration as time passed by taking sequences of snapshots. Quantitatively, we determined the time-dependent disorder parameter and structure factors.

From the numerical data analysis of the slope of $A(L,T;t)$ vs. $t$, we found that the scaling relation no longer held, especially for low $T$, due to the fact that inter-particle interactions had begun to play a role. Correlation in the system switches from being short-range, so that the vacancy does not perform a random walk. We fitted the slope of the log-log plot of the disorder parameter versus time for various temperatures by using an exponential and a Gompertz function. The later seemed to give a better fit.

We now present some observation of the structure factors. The dips seen before are still present, but become more and more obscured as the temperature is lowered. The portion of the graph exhibiting a scaling exponent of slope = 1 in the 'I' region now is replaced earlier by a graph having a slope = 0.5. In conclusion, both measurements suggest that the correlations due to the finite temperature may break the universality of the scaling seen at infinite $T$. This is caused by the microscopic mechanism of how vacancies move and how the non-linearities interact with the environment. Even though the model we considered is a simple one, it can be the basis for describing of a large variety of related systems.

## ACKNOWLEDGMENTS

This research is supported in part by the Thailand Research Fund through grant numbers TRG4580090 and RTA4580005 and by MTEC Young Research Group funding (MT-NS-45-POL-14-06-G). The Royal Golden Jubilee Ph.D. Program (PHD/0240/2545) to Jirasak Wong-ekkabut and I-Ming Tang is acknowledged.